\documentstyle[preprint,aps,12pt]{revtex}

\begin{document}

\tightenlines
\draft

\title{Anomalous structure factor of dense star polymer solutions}

\author{M Watzlawek\dag, H L{\"o}wen{\dag\ddag} and C N Likos\ddag}
\address{\dag Institut f{\"u}r Theoretische Physik II,
   Heinrich-Heine-Universit{\"a}t D{\"u}sseldorf,\\
   Universit{\"a}tsstra{\ss}e 1, D-40225 D{\"u}sseldorf, Germany\\
   \ddag Institut f{\"u}r Festk{\"o}rperforschung, Forschungszentrum J{\"u}lich
   GmbH, D-52425 J{\"u}lich, Germany}

\date{June 12, 1998}

\maketitle

\begin{abstract}
   The core-core structure factor of dense star polymer solutions in a good
   solvent is shown theoretically to exhibit an unusual behaviour above
   the overlap concentration. Unlike usual liquids, these solutions
   display a structure factor whose first peak decreases by increasing
   density while the second peak grows. The scenario repeats itself with
   the subsequent peaks as the density is further enhanced. For low enough
   arm numbers $f$ ($f \leq 32$), various different considerations lead to
   the conclusion that the system remains fluid at all concentrations.
\end{abstract}


\narrowtext

\section{Introduction}

Star polymers are macromolecular entities consisting of $f$ polymeric
arms chemically attached to a common centre. At the limit where the
number of monomers per arm (degree of polymerisation) $N$ is large,
the size of the central core of the star is negligible compared
with the overall radius of the macromolecular aggregate and can be
ignored at a first approximation. Scaling theory has provided
insight into the structure and conformation of a
{\it single} star and the way that these properties are influenced
by the arm number (functionality) $f$, $N$ and solvent quality
\cite{daoud,reviewsp}. From both theoretical and experimental points of
view, it is even more interesting to describe the properties of
{\it dense} star polymer solutions by means of an effective pair
potential between star centres. Experimentally, this is important
as interstar correlations are ubiquitous in understanding neutron
scattering spectra of dense star polymer solutions. Moreover,
such a description of stars establishes a bridge between
polymer and colloidal science: indeed, due to their peculiar construction,
star polymers can be viewed as hybrids between polymers and colloidal
particles. Once such a description has been established, it is then
possible to look at a star polymer solution as an effective one-component
system of point particles (the star centres) and employ the known
machinery from liquid state theory and/or computer simulations to
study its properties.

By way of direct comparison with experimental data, it was found recently
that star polymers in a good solvent can be described by an effective
pair potential which is logarithmic for short distances and crosses
over to a Yukawa form for larger interstar separations \cite{likos}.
This is a new type of interaction in the sense that (i) it has an ultra-soft
logarithmic core whose hardness depends in fact on the functionality $f$
in a way which will be explained shortly and (ii) it features a
crossover from one functional form to another at some length scale
$\sigma$ which is of the order of the corona diameter of the star.
Systems which are described by simple, spherically symmetric interactions
have been studied extensively in the last thirty years, since the
development of powerful computers made integral equation theories
and simulations computationally tractable \cite{hansen,allen,frenkelsmit}.
For most
of the commonly considered interactions (power laws, Yukawa, square-well
or square-shoulder potentials etc.) the liquid structure factor $S(k)$
features a single characteristic length which is basically set by
the density. Moreover, the correlations in structure grow
with increasing density $\rho$
until, at some temperature-dependent value, the system crystallises.
We call the liquids for which such a scenario materialises {\it usual}
or {\it normal}. The purpose of this paper is to show that star
polymer solutions are unusual, in the sense that $S(k)$ or its
real-space counterpart, the radial distribution function $g(r)$
have quite unexpected behaviour above the overlap concentration of
the solution. The latter is defined as the polymer concentration
at which the stars start overlapping within their corona.
There exist two competing length scales which manifest
themselves in the form of a structure factor
whose peaks do not grow simultaneously
in height as $\rho$ increases, but rather lower order peaks grow higher
until they completely dominate and the original main peak disappears.
These two lengths are the average interparticle distance $a \propto
\rho^{-1/3}$ and the corona diameter $\sigma$ of the stars.
Moreover, for low enough functionality, freezing does not take place,
i.e., the system remains fluid at all densities.

The rest of the paper is organised as follows: in section 2 we present
the model pair potential and briefly discuss its properties. In section 3
we present results from computer simulations and integral equation
theories regarding the structure of the system for a very wide range
of densities demonstrating the unusual features and we discuss their
origin. In section 4 we present a simple
`Hard-Sphere mapping' argument to establish the limit of $f$ beyond
which the system does not crystallise. In section 5 we discuss the
connection with experiments.
Finally, in section 6 we summarise and conclude.

\section{Pair potential between star polymers}

Our starting point is the pair potential
between two star polymers in
a good solvent separated by a centre-to-centre distance $r$,  which reads as:
\begin{eqnarray}
{V(r)\over{k_BT}}=\cases{(5/18) f^{3/2}
     \Bigl[-\ln(r/\sigma) + (1+\sqrt{f}/2)^{-1}\Bigr]
     \;\; (r \leq \sigma); \cr
            (5/18) f^{3/2} (1+\sqrt{f}/2)^{-1}
            (\sigma/r) \exp[-\sqrt{f}(r-\sigma)/2\sigma]
     \;\;(r > \sigma),}
\label{poten}
\end{eqnarray}
where $k_B$ is Boltzmann's constant and $T$ is the absolute temperature.
This is an entropic interaction stemming from excluded volume effects
between monomers belonging to two different stars in a good solvent.
It was shown recently \cite{likos}
that the length scale $\sigma$ has to be chosen as twice the distance
between the centre of the star and the centre of the largest (outermost)
blob \cite{footnote}.
The logarithmic form of the interaction for interparticle distances
$r < \sigma$ results from the arguments of Witten and Pincus \cite{witten}.
For $r > \sigma$ we take an
exponential form for the interaction with a decay
length equal to the largest blob diameter. For details on
the determination of the overall numerical prefactor
of the logarithmic term, we refer the reader
to Ref.\ \onlinecite{likos}. The amplitude of the Yukawa tail is finally
determined by the requirement of smoothness of the interaction at
$r = \sigma$. In what follows we use $\sigma$ as the unit of length
and introduce a dimensionless `packing fraction' $\eta$ defined as:
\begin{eqnarray}
   \eta \equiv {{\pi}\over{6}}\rho\sigma^3.
   \label{packfr}
\end{eqnarray}

It has been shown in
Ref.\ \onlinecite{likos} that this choice of a pair potential \cite{foot1}
brings about good agreement
between theory and experiment regarding small-angle neutron scattering
(SANS) data for a wide range of densities (or polymer volume fraction).
Moreover, a similar exponential `cutoff' of the logarithmic part of the
interaction has also been employed recently in a study of the elastic
moduli of block copolymer micelles \cite{foerster}. The entropic nature
of the interaction renders the temperature $T$ irrelevant, as the
Boltzmann factor $e^{-\beta V(r)}$ is temperature-independent.
Instead, the arm number $f$ plays now the role of an effective
inverse temperature. Indeed, in the limit $f \to 0$ the interaction
vanishes, whereas at the ``colloidal limit'' $f \to \infty$ we
recover the well-known hard sphere (HS) interaction.
In fig.\ \ref{stars} we show the pair potential for different
values of $f$. Whereas for large $f = 128$ and $256$ there is a
spectacular change in its behaviour as $r$ crosses $\sigma$, for
low $f$-values, the potential is ultra-soft and its range becomes
longer.

As will be shown in section 4, the ultra-soft character of the
interaction at hand, 
has the consequence that, for low enough values of the functionality $f$,
the system never crystallises, no matter how
large the density or external pressure are. Thus, this system offers
us a unique possibility to examine the behaviour of the structural
functions of the liquid (the radial distribution function and/or
the structure factor) over a practically unlimited range of densities.
This is the subject of the following section.

\section{Anomalous structure factor}

First, we summarise some basic notions from the theory of classical
liquids and refer the reader to Ref.\ \onlinecite{hansen} for details.
A quantity of central interest for a classical fluid in equilibrium
is the so-called radial distribution function $g(r)$ and the closely
associated pair correlation function $h(r) \equiv g(r) - 1$. If
we call $\rho$ the number density of the liquid, then the
quantity $\rho g(r)$ is nothing else but the density profile that
develops if we keep a particle fixed at the origin. In other words,
the quantity $g(r)$ expresses the ordering of the rest of the system
around a given particle of the liquid. Equivalently, one can look
at the structure factor $S(k)$ which is simply $\rho$ times the
Fourier transform of the pair correlation function. The peaks of
$S(k)$ reveal information about the characteristic length scales
in the system. Alternatively, $g(r)$ is more appropriate if one
wishes to determine, e.g., the average coordination number of the
liquid.

Both $g(r)$ and $S(k)$ can be measured in a standard Monte Carlo
simulation \cite{allen,frenkelsmit}.
We have thus performed such simulations
for a wide range of densities and, in addition, we have solved the
Rogers-Young (RY) closure \cite{ry} in order to obtain a comparison.
After ascertaining that the RY gives quite reliable results for
all the densities for which simulations were carried out, we relied
on this closure to calculate the structure of the fluid for very
high packings
($ 60.00 {> \atop \sim} \eta {> \atop \sim} 10.00$),
where a simulation would be very
expensive since a very large number of particles
in the simulation box would be necessary in order to obtain
reliable results.

In fig.\ \ref{gofrsim} we show representative results in order to
provide a comparison of the radial distribution
function $g(r)$ obtained from simulations with the RY result for
various different values of $\eta$. Apart from small discrepancies
for the intermediate value $\eta = 0.60$, it can be seen that the
agreement is quite good. Hence, the RY closure is a reliable theoretical
tool for the calculation of the pair structure of the liquid. It should
also be noted that for high values of $\eta$, $\eta {> \atop \sim} 3.0$,
the RY closure reduces practically to the Hypernetted Chain (HNC) for
our system. This is expected since, for high densities, we are dealing
with a long-range interaction when distances are measures in units
of average interparticle distance and the HNC is known to be most
accurate precisely for long-range potentials.

The radial distribution function $g(r)$ of the system at hand, shows
the typical evolution of a normal liquid as the density is increased
until we reach the overlap density $\rho^{*}$.
This is 
the density at which the average interparticle
distance $a = \rho^{-1/3}$ becomes equal to the length scale $\sigma$.
It corresponds roughly to an overlap packing faction $\eta^{*} = 0.50$.
For $\eta < \eta^{*}$, the function $g(r)$ displays oscillations with
a characteristic length scale $a$. The heights of the peaks of $g(r)$ 
grow with
density; the same behaviour is also observed for the structure
factor $S(k)$, of course.

The situation changes above the overlap concentration, as can be seen
in fig.\ \ref{gofranom}(a). Instead of having a main peak of $g(r)$
at $r \approx a$, we observe that one peak of $g(r)$ always remains
at slightly outside the logarithmic part at $r {> \atop \sim} \sigma$
{\it regardless of the density}. The `main' peak of $g(r)$ grows
gradually as the density is increased. Indeed, at $\eta = 0.80$
$g(r)$ shows just a shoulder preceding the peak at $r {> \atop \sim} \sigma$
mentioned above and only at $\eta \geq 1.00$ can one distinguish a
clear peak at $r \approx a$. Even then, this peak remains lower
than the second until $\eta = 1.30$. Thereafter, it overpasses the
second peak in height but still, the position of the second
peak of $g(r)$ is not at twice the distance of the first peak,
as is the case for normal liquids, because the latter is determined
by the density and the former by the length scale $\sigma$.

This anomaly in $g(r)$ is reflected, naturally, in the shape and
evolution of the structure factor $S(k)$, as can be seen in
fig.\ \ref{gofranom}(b). As a first remark, we find that the height
of the main peak of $S(k)$, which grows until we reach the
overlap packing, becomes {\it lower} above $\eta^{*}$. More unusual
features are observed if one looks at the positions and the
competition between the first two peaks. Though the second peak
remains at twice the position of the first up to $\eta = 0.60$, at
packing $\eta = 0.80$ the position of the first peak is practically
unchanged with respect to that at $\eta = 0.60$, but the second
peak now {\it moves closer} to the origin and increases in height.
This trend persists as $\eta$ grows and already at $\eta = 1.30$ the
second peak has become higher than the first, which shows a clear trend
of disappearing at higher densities, as it will be confirmed shortly.

This anomaly in the structure factor can be traced to the crossover of
the interaction from a Yukawa to a logarithmic form {\it and} the
softness of the logarithmic potential. Indeed, above the overlap
concentration the system always tries to maintain one coordination
shell {\it outside} the logarithmic part, where the interaction is
weak. At the same time, the softness of the logarithmic core allows
two things to happen: on the one hand, qualitatively, the system
remains always fluid which would not have been the case if the core
was harder (e.g. a HS core, or even the same, logarithmic form for
different $f$, as will be shown in section 4.) On the other hand,
more quantitatively, the softness of the logarithm allows for a
rather broad first peak of the radial distribution function which can
accommodate enough particles so as to allow for the second peak to
occur immediately outside $r = \sigma$.

The unusual shape of the structure
factor can be understood by means of the existence in $g(r)$ of
these two different length scales: one length scale $a \approx \rho^{-1/3}$,
which is manifestly density-dependent and one length scale
$b {> \atop \sim} \sigma$ which is density-independent. Below $\eta^{*}$
only the first length scale appears but above $\eta^{*}$ both are
present. Let us call $k_n$ the position of the $n$-th peak of the
structure factor.
The first peak of $S(k)$ corresponds, roughly, to the
length scale $b$, i.e.\ $k_1 \approx 2\pi/b$ and the second peak to the
length scale $b-a$, i.e.\ $k_2 \approx 2\pi/(b-a)$. Indeed, as can be seen
from fig.\ \ref{gofranom}(b), $k_1$ is practically constant, whereas
$k_2$ decreases with density, a feature that can be attributed to the
increase of the quantity $b-a \approx \sigma - \rho^{-1/3}$ with growing
density. Moreover, the growth of the second peak can be understood since
the structure that gives rise to it becomes more pronounced with
increasing $\eta$, as the first peak of $g(r)$ takes shape.

In these terms, we can now make the hypothesis that the first peak of the
structure factor will disappear altogether when the density is such that
$b-a = a$ or $b = 2 a$. In other words, the structure factor will
have a main peak at a position dictated only by the density, when
the length $b$ is twice the length $a$ in such a way that $g(r)$ has
exactly two oscillations of wavelength $a$ between $r = 0$ and $r = b$.
Since $b \approx \sigma$, it turns out that the density must be such
that $a \approx \sigma/2$.
As can be seen in fig.\ \ref{eta2}(a) this occurs at the `magic' value
$\eta_2 = 3.40$, where the subscript denotes the number of oscillations
of $g(r)$ in the interval $[0, b]$. Accordingly, the structure factor
at $\eta = \eta_2$ has a strong first peak (which, however, has evolved
from the second peak at lower densities!) located at a position
$2\pi/a \approx 4 \pi/\sigma$.
In fig.\ \ref{eta2}(b) we show also the structure
factor at $\eta = 2.00$ in order to show the disappearence of what
used to be the first peak of $S(k)$ and its replacement by the
second one. However, we emphasize that the fact that at $\eta = \eta_2$
we have again a structure factor with its main peak located at a position
$k_{max} \approx \rho^{1/3}$, does not mean that we are dealing with
a normal liquid. Indeed, as can be seen from fig.\ \ref{eta2}(b) the
structure factor shows some weak substructure which is not present for
usual liquids.

It is now pertinent to ask the question what happens once the density
is further increased. Does the height of the new main peak
grow until the system solidifies or does the above scenario repeat
itself? We have solved the RY-closure for packing fractions up to
$\eta = 60.00$ and we find that, in fact, it the second possibility
that materialises. As $\eta$ grows, the main peak is lowered and the one
that used to be the third
at low densities grows, see fig.\ \ref{sofk}(a).
The mechanism that brings about this scenario
is no other but the development of more and more oscillations
inside the logarithmic core in the function $g(r)$, as can be seen
in fig.\ \ref{sofk}(b). In fact, one can repeat the argument about
the `magic' packing fraction $\eta_2$ above for an arbitrary number
of oscillations as follows: the structure factor $S(k)$ will be
dominated by a single length scale whenever there is an integer
number of oscillations $m$ between $r = 0$ and $r = b \approx \sigma$.
Given the magic packing fraction $\eta_2$, it is straightforward to show
that the general magic packing fraction $\eta_m$ will
be related to $\eta_2$ by:
\begin{eqnarray}
   \eta_m = \Bigl({{m}\over{2}}\Bigr)^{3} \eta_2, \qquad\qquad m = 3,4,5,\cdots
\label{magetas}
\end{eqnarray}

Using $\eta_2 = 3.40$ we obtain $\eta_3 = 11.475$, $\eta_4 = 27.20$ and
$\eta_5 = 53.125$. The structure factor for these magic values is
shown in fig.\ \ref{multip}. It can be seen clearly that for the $m$-th
magic value, $S(k)$ has a dominant peak located at the position
$2 \pi m/\sigma$. As the order $m$ grows, the height of the dominant peak
also decreases slightly and some substructure develops in $S(k)$.
However, the main length scale comes from the $m$ oscillations of the
radial distribution function $g(r)$ in the logarithmic core.

The above results can be nicely summarised by making a log-log plot
of the positions of the first few peaks of the structure factor
against the packing fraction $\eta$, as shown in fig.\ \ref{loglog}.
We emphasise here that when we talk about the $n$-th peak we actually
mean `the peak which is the $n$-th if we extrapolate at low enough
densities so that we are in a regime where the liquid is normal.'
The reason for this distinction is that as $\eta$ grows the low-order
maxima of the structure factor disappear, as explained above,
and as a result what used to be a second-order maximum becomes
now first order etc. This is manifested in fig.\ \ref{loglog} by
the fact that the curves representing the positions of the various
maxima $k_n$ stop at some value of $\eta$. Moreover, since a higher-order
peak overtakes a lower-order one in height before the latter completely
disappears, we indicate the highest peak in fig.\ \ref{loglog} by
the filled symbols.

Referring to fig.\ \ref{loglog}, we can make the following remarks:
according to our previous definition, the fluid interacting by means
of the potential given by eq.\ (\ref{poten}) is normal for packing
fractions $\eta \leq \eta^{*} = 0.50$. Indeed, in this regime on
the one hand the first maxima follow the scaling $k_1 \propto \rho^{1/3}$
and on the other hand the higher-order maxima $k_n$ are located in
positions $k_n = n k_1$, both features being a manifestation of the
existence of a {\it single} length scale $\rho^{-1/3}$ in the
structure of the system. Above the overlap density $\rho^{*}$
the scaling breaks down. However, if we extrapolate the straight line
with slope $1/3$ which characterises the normal regime to higher
densities, we find that it passes precisely through the main maxima
at the magic packing fractions $\eta_m$. Indeed, at those values of
the density the length scales $\rho^{-1/3}$ and $\sigma$ are
commensurate and we have an accidental scaling of the main peak of
$S(k)$ with the $1/3$-power of the density. However, the higher-order
peaks of $S(k)$ are still not located at integer multiples of the
first one and we are dealing with an unusual fluid for all densities
exceeding the overlap density $\rho^{*}$.

\section{Hard sphere mapping and the freezing transition}

The structure factors obtained for the whole range of densities
have the characteristic that the maximum height of the main peak
in $S(k)$ never exceeds the value 2.8. According to the empirical
Hansen-Verlet criterion \cite{hanver}, a liquid freezes when the
first maximum of the structure factor reaches a value 2.85. It was
subsequently shown that indeed freezing sets in if the order in the
liquid phase, as measured by the first maximum of $S(k)$ exceeds
this quasi-universal value \cite{hansenschiff,hartmut}.
Hence, we have a first
indication that for $f = 32$ the system does not freeze. Naturally,
the same must also hold for smaller $f$-values as the logarithmic
core is then even softer. The absence of freezing was also
confirmed in our numerical simulations, where for all values of
$\eta$ which were simulated ($\eta \leq 12.00$) the system remained
in a liquid-like configuration, with any order parameter
of a hypothetical crystalline solid vanishing.

Here, we want to apply a simple Hard-Sphere mapping procedure in order to
corroborate the above results for $f = 32$ on the one hand and establish
approximately the `critical arm number' $f_c$ on the other,
which is defined as
follows: for $f\leq f_c$ the system remains fluid at all densities,
but for $f>f_c$ there is at least one region in the density domain
where the system is crystalline.

Following an idea of Kang {\it et al.} \cite{kang}, we define an
effective Hard-Sphere diameter $\sigma_{HS}$, crudely representing
the particle repulsion embodied in the pair potential of 
eq.\ (\ref{poten}), as follows.
The pair potential $V(r)$ is divided into a short-ranged reference
potential $V_{0}(r)$ and a longer-ranged
perturbation potential $W(r)$ at a suitably chosen break point
$\lambda$. Thus, we write
\begin{eqnarray}
V(r)=V_{0}(r)+W(r),
\label{V.split}
\end{eqnarray}
where $V_{0}(r)$ is given by
\begin{eqnarray}
V_{0}(r)=\Theta(\lambda-r)\left[V(r)-F(r)\right].
\label{V.reference}
\end{eqnarray}
Here, $\Theta(r)$ is the Heaviside step function, and $F(r)$ will be
defined soon.
According to eqs.\ (\ref{V.split}) and (\ref{V.reference}), the
perturbation potential $W(r)$ is given by:
\begin{eqnarray} 
W(r)=\Theta(\lambda-r)\, F(r)+ \Theta(r-\lambda)\, V(r).
\label{V.perturbation}
\end{eqnarray}
Having splitted the potential in this way, it is now
possible to calculate the free energy of the system
by a simple Hard-Sphere perturbation theory \cite{kang,lutsko}.
In this theory, the potential
$V_{0}(r)$ is used to calculate an effective Hard-Sphere diameter
$\sigma_{HS}$, while the longer-ranged potential $W(r)$
is treated in a mean-field approach. In the following,
we are not going to
trace out calculations of free energies, but only to calculate
$\sigma_{HS}$, which is density-dependent when the specific
choice for $\lambda$ and $F(r)$, applied successfully
in Refs.\ \cite{kang,lutsko}, is used. This density-dependence
arises from the identification of $\lambda$
with the nearest-neighbour distance $a_{fcc}$ of the $fcc$ structure, i.e.,
\begin{eqnarray} 
\lambda=a_{fcc}\equiv\left(\frac{\sqrt{2}}{\rho}\right)^{1/3}.
\label{lambda.fcc}
\end{eqnarray}
Furthermore, the function $F(r)$ is chosen as
\begin{eqnarray} 
   F(r)=V(\lambda)-\left[\frac{dV(r)}{dr}\right]_{r=\lambda}
                   \left(\lambda-r\right),
\label{F.potential}
\end{eqnarray}
guaranteeing that both $V_{0}(r)$ and $dV_{0}(r)/dr$ are vanishing
at $r=\lambda$.
Having specified $\lambda$ and $F(r)$, $\sigma_{HS}$
is then calculated from $V_{0}(r)$ by the well-known
Barker-Henderson (BH) approximation \cite{barkerhenderson}
\begin{eqnarray} 
   \sigma_{HS}=\int_{0}^{\infty}~dr~
               \left[1-e^{-\beta V_{0}(r)}\right],
\label{sigma.HS}
\end{eqnarray}
or, alternatively, by a scheme proposed by Weeks, Chandler and
Anderson (WCA) \cite{kang,lutsko,WCA}. We
will only present results from the BH approximation, since the
corresponding WCA results are practically identical to the latter,
thus leading to the same conclusions.

After obtaining $\sigma_{HS}$ from the BH or WCA scheme,
we calculated the effective Hard-Sphere packing fraction
$\eta_{HS}\equiv\frac{\pi}{6}\rho\sigma_{HS}^{3}$,
which is in general density-dependent. We can hence plot
$\eta_{HS}$ as a function of the `true' packing fraction $\eta$,
depending furthermore on the value of $f$ as additional parameter.
In fig. \ref{etahs}, we show results for $\eta_{HS}$ for five
different functionalities $f$.
Obviously, $\eta_{HS}$ is increasing linearly with $\eta$ for
$\eta{< \atop\sim}0.1$, and it reaches a maximum value depending
on $f$ at certain packing fractions
$0.4{<\atop\sim}\eta{<\atop\sim}0.7$, which are again
depending on $f$. For $\eta\geq 0.74$, $\eta_{HS}$ remains constant
for all densities.

We will now briefly explain this behaviour of
$\eta_{HS}(\eta)$, before switching to some conclusions that can 
be drawn from
fig.\ \ref{etahs} concerning the freezing transition of star polymers.
For small $\eta$, where the break point $\lambda$ is located at
distances larger than the range of the pair potential, eqs.\
(\ref{V.reference}), (\ref{F.potential}) and (\ref{sigma.HS}) lead
to a density-independent $\sigma_{HS}$,
since the pair potential $V(r)$ is density-independent.
Hence, $\eta_{HS}$ scales linearly with $\eta$ in
this regime. However, when $\lambda$ reaches distances where
the potential is remarkably different from zero, this linear
scaling is no longer valid, since, according to eqs.\
(\ref{V.reference}) and (\ref{sigma.HS}),
$\sigma_{HS}$ is now a decreasing function with increasing
density. This fact materialises in a decreasing slope of
the function $\eta_{HS}(\eta)$, leading even to a existence
of a maximum in $\eta_{HS}(\eta)$. Having these arguments in mind,
the surprising fact that $\eta_{HS}$ attains a constant value
for $\eta\ge\eta_{c}=0.74$ can be explained as follows. Exactly
at the cross-over packing fraction $\eta_{c}$, the
break point $\lambda$ is located at the corona diameter $\sigma$ and
thereafter, for $\eta > \eta_c$ we have $\lambda < \sigma$. 
Therefore, for $\eta \geq \eta_c$
the Yukawa part of the interaction potential $V(r)$
is irrelevant for the calculation of
$\sigma_{HS}$, as the reference potential $V_0(r)$ is purely logarithmic.
Using eqs.\ (\ref{poten}) and (\ref{sigma.HS}), it follows that in this
regime $\sigma_{HS}$ scales linearly with $\lambda$, i.e.,
\begin{eqnarray}
   \sigma_{HS}={\cal A}(f)~\lambda,
\label{sigma.hs.lambda}
\end{eqnarray}
with the constant ${\cal A}(f)$ depending only on the
arm number $f$, but neither on $\sigma$ nor on $\lambda$.
Since $\lambda$ scales with $\rho^{-1/3}$, 
eq.\ (\ref{sigma.hs.lambda}) leads directly to the observed
density-independence of $\eta_{HS}(\eta)$ 
for $\eta\ge\eta_{c}$ \cite{bccnote}.

The logarithmic pair
interaction at hand is the only one showing this feature. Indeed, 
in order to have $\sigma_{HS} \propto \lambda$, the integrand in 
eq.\ (\ref{sigma.HS}) must be a function of $r/\lambda$ {\it only},
i.e., the length scale $\sigma$ must drop out of the expression
for $V_0(r)$. Let us assume that the pair potential $V(r)$ is given by
some function $R(r/\sigma;f)$ for $r \leq \lambda$.
If the reference potential 
$V_0(r)$ has to depend on $r/\lambda$ only, the function $R(x;f)$ must 
satisfy the following relations,
as is clear from eqs.\ (\ref{V.reference}) and
(\ref{F.potential}) above:
\begin{eqnarray}
R(r/\sigma;f) - R(\lambda/\sigma;f) = \bar R(r/\lambda;f) 
\qquad {\rm and}
\qquad {{dR(r/\sigma;f)}\over{dr}} \propto {{1}\over{r}},
\label{conds}
\end{eqnarray}
where $\bar R(x;f)$ is some other function. 
The above conditions (which are, in fact, not independent but equivalent
to each other) are fulfilled only by the family of functions
$R(x;f) = C(f) \ln(x) + D(f)$,
with $C(f)$ and $D(f)$ arbitrary.

We now turn to the main reason for our
interest in the values of the effective Hard-Sphere packing fraction
$\eta_{HS}$. Since the mapping onto an effective Hard-Sphere
system gives quite reliable
results in predicting the freezing transition \cite{kang,lutsko},
we use the value of $\eta_{HS}$
as an indication for the existence of a freezing transition in the
original system. Hard Spheres freeze in a $fcc$ structure
at $\eta_{HS}^{solid}=0.55$ \cite{bolhuis},
and we therefore take this specific
value to explore a possible freezing transition
of star polymers from fig. \ref{etahs}.
Obviously, for $f\le 32$, $\eta_{HS}$ never
exceeds $0.55$ for all $\eta$, leading to the
conclusion that the system remains fluid
at {\em all densities}. For $f>32$ on the other hand,
$\eta_{HS}$ attains values larger than $0.55$ at least in a limited
window of packing fractions $\eta$. Consequently,
there is a critical arm number $f_{c}\simeq 32$, meaning that the
system never freezes for $f\le f_{c}$, but freezes presumably
at least in a limited range of densities for $f>f_{c}$.
Our previously described finding that star polymers with $f=32$
did not freeze in all our computer simulations is therefore
consistent with this crude Hard-Sphere mapping
procedure.

As can be further seen from fig. \ref{etahs}, there is a range of arm numbers
$f_c<f{<\atop\sim}64$, where $\eta_{HS}$ exceeds $0.55$ only for
$0.2{<\atop\sim}\eta{<\atop\sim}0.7$, which implies
a `reentrant-melting'-phenomenon at $\eta\simeq 0.7$, i.e., a transition
form a solid phase to a liquid phase if $\eta$ is increased above $0.7$.
Here, it is worth mentioning that reentrant melting for star polymer
solutions at high concentrations was predicted by Witten {\it et al.}
on the basis of arguments arising from scaling theory \cite{cates}.
This prediction is independently verified here. Moreover,
within our crude model, star polymers with $f{>\atop\sim}64$
show a liquid phase for $\eta{<\atop\sim}0.2$, followed by a solid phase for
{\em all} higher densities.

\section{Connection with experiments}

From the experimental point of view, the extreme values of the
packing fraction that we have considered in section 3 are unattainable.
At most, one can expect to observe a change in the behaviour of
the first peak of the structure factor as the overlap polymer
concentration (which corresponds to our overlap density) is
crossed. Our prediction is that the height
of the first peak of the star-star
structure factor of a star polymer solution in a good solvent
is not monotonically increasing with polymer concentration but
rather it saturates at the overlap density and decreases thereafter.
Based on general scaling properties of polymers in good solvents,
Witten {\it et al.} predicted, more than ten years ago \cite{cates},
precisely that the peak of $S(k)$ is largest when the separation
between stars is of the order of the star radius. Here, we have
confirmed this prediction quantitatively by employing a
{\it colloidal} description of star polymer solutions.
We expect that such effects should be visible in SANS experiments
of dense star polymer solutions. To this end, stars with a labeled
core should be used and the extent of the labeled
part should be made as small as possible. This way, the effects
of increasing concentration on the structure factor will not be
masked by the form factor of the single star.

Additionally, it is natural to ask whether this nonmonotonic behaviour
in the peak of $S(k)$ is peculiar to the logarithmic form of the
interaction inside the core or it can be seen for other functional
forms of the pair potential as well. To test this, we have taken
a toy potential which has a soft core for separations smaller
than some length $\sigma$ and a crossover to a different functional
form for larger
separations. This toy potential reads as:
\begin{eqnarray}
{\phi(r)\over{k_BT}}
      =\cases{A (r-\sigma)^2/(r\sigma) \;\; (r \leq \sigma); \cr
                     0 \;\;(r > \sigma),}
\label{toy}
\end{eqnarray}
where $A$ is some numerical constant which we can vary and controls the
steepness of the interaction inside the core. We take now $A = 10$.
Defining the packing fraction $\eta$ as in eq.\ (\ref{packfr}) above,
we have solved the HNC closure for a few different $\eta$-values.
The results are shown in fig.\ \ref{toysofk}, where it can be seen
that the same nonmonotonic behaviour of the main peak of $S(k)$ is
observed. Thus, the phenomenon is rather general and it relies on
the existence of a soft enough core in the interaction. Such 
interactions are not uncommon in soft-matter physics. Hence, it would
be of great interest if such anomalies could be experimentally
observed.

We finally comment on similarities of the logarithmic potential
investigated in this work to a model introduced 
by Uhlenbeck and Ford in the early days of liquid-state theory 
\cite{uhlenbeck}. 
This special model is defined by a Gaussian 
Mayer-function $f(r)=1-\exp(-\beta V(r))=\exp(-\alpha r^{2})$, 
corresonding to the pair potential 
$
   \beta V(r)=-\ln(1-\exp(-\alpha r^{2})).
$
As shown in Ref. \cite{uhlenbeck}, 
all interparticle correlations can be calculated analytically 
within the framework of graph theory.
For small distances $r$, the above potential potential
reduces to $\beta V(r)\simeq -2\ln(\sqrt{\alpha}r)$, corresponding to 
the potential of eq. (\ref{poten}) with an 
arm number $f=(36/5)^{2/3}\simeq 3.7$.
Since this arm number is significantly smaller than $f_{c}$, we
expect the model system of Uhlenbeck and Ford to be liquid for all
densities, corresponding to an equation of state analytical in the 
density.

\section{Summary and concluding remarks}

Employing a pair potential which has been shown to describe correctly
the SANS data of star polymer solutions in good solvents \cite{likos},
we made quantitative predictions regarding the behaviour of the 
structure factor as a function of increasing polymer concentration.
In particular, we found that above the overlap concentration, star
polymer solutions display features which are unknown for normal
liquids, namely a breakdown of the $\rho^{1/3}$-scaling of the 
positions of the peaks of $S(k)$ as well as an anomalous evolution
of the heights of these peaks: the lower-order peaks diminish in
height and the higher-order ones grow. 

Furthermore, we applied a Hard-Sphere mapping and used it as a crude
diagnostic tool in order to make preliminary investigations on the
topology of the phase diagram of star polymer solutions.
The most striking feature of the Hard-Sphere mapping procedure is,
in our view, the fact that the effective HS packing fraction $\eta_{HS}$
remains constant at high densities and depends only on the functionality
$f$. We have shown that this characteristic is particular to the 
logarithmic interaction. We speculate that interactions which are
softer than the logarithm will lead to an $\eta_{HS}$ which will be
a {\it decreasing} function of the density. This has relevance, in
particular, to `bounded' interactions, i.e., interactions which do not
diverge at the origin, such as the Gaussian potential of
Stillinger \cite{still} or a simple model of penetrable spheres
introduced recently by us \cite{pen}. 

Another question of great interest is the phase diagram of star
polymers. Detailed
calculations, based on extensive computer simulations
as well as perturbation theory
are currently under way. These calculations allow for the evaluation
of the Helmholtz free energies of various candidate crystalline phases
and, subsequently, the comparison of the latter with that of the 
fluid phase and the construction of the phase diagram. Preliminary
results are in full agreement with those presented here as far as 
the critical arm number $f_c$ is concerned; at the same time, they
reveal a rich topology of the phase diagram as well as a variety of
unusual crystal structures. The presentation of the phase diagram of
star polymers will be the subject of a future publication.

\section*{Acknowledgments} MW thanks the Deutsche Forschungsgemeinschaft
for support within the SFB 237.

\begin{figure}
\caption[dum01] {The pair potential given by Eq.\ (\ref{poten}) for $f=18,
32, 64, 128$, and $256$ as a function of the
centre-to-centre separation $r$.}

\label{stars}
\end{figure}

\begin{figure}
\caption[dum02] {The radial distribution function $g(r)$ as obtained
from simulations and from the RY closure for different $\eta$-values.
(a) $\eta = 0.10$; (b) $\eta = 0.60$; (c) $\eta = 1.50$.}

\label{gofrsim}
\end{figure}

\begin{figure}
\caption[dum03] {(a) The radial distribution function $g(r)$
and (b) the anomalous structure factor $S(k)$ above the overlap
density.}

\label{gofranom}
\end{figure}

\begin{figure}
\caption[dum04] {(a) The radial distribution function $g(r)$
and (b) the structure factor $S(k)$ as obtained from simulations
and from the RY closure at $\eta = \eta_2 = 3.40$.
For comparison, also the structure factor at $\eta = 2.00$ is shown.}

\label{eta2}
\end{figure}

\begin{figure}
\caption[dum05] {(a) The structure factor $S(k)$ and (b) the radial
distribution function $g(r)$ for values of the packing fraction in
the interval $\eta_2 < \eta < \eta_3$.}

\label{sofk}
\end{figure}

\begin{figure}
\caption[dum06] {The structure factor $S(k)$ at the magic
values of the packing fraction
$\eta_m$, for $m = 2$, 3, 4 and 5.}

\label{multip}
\end{figure}

\begin{figure}
\caption[dum07] {Log-log plot of the positions of the various peaks
of $S(k)$ against the packing fraction $\eta$. The filled symbols
indicate the highest peak (see the text). The arrows
indicate the locations of the magic packing fractions $\eta_m$
and the dashed straight line has slope 1/3.}

\label{loglog}
\end{figure}

\begin{figure}
\caption[dum08] {The effective Hard-Sphere packing fraction $\eta_{HS}$
obtained by the procedure described in section 4 as a function of the true
packing fraction $\eta$ for five different values of
the functionality $f$.}

\label{etahs}
\end{figure}

\begin{figure}
\caption[dum09] {The structure factor of a system interacting by means of the
potential given by eq.\ (\ref{toy}) with $A=10$
at various packing fractions $\eta$,
obtained in the HNC closure. Notice the nonmonotonic behaviour of the
maximum of $S(k)$ with increasing density.}

\label{toysofk}
\end{figure}


\begin{references}

\bibitem{daoud} Daoud M and Cotton J P 1982 {\it J. Physique} {\bf 43} 531

\bibitem{reviewsp} Grest G S, Fetters L J, Huang J S and Richter D 1996
{\it Adv. Chem. Phys.} {\bf XCIV} 67

\bibitem{likos} Likos C N, L{\"o}wen H, Watzlawek M, Abbas B, Jucknischke O,
Allgaier J and Richter D 1998 {\it Phys. Rev. Lett.} {\bf 80} 4450

\bibitem{hansen} Hansen J P and McDonald I R 1986 {\it Theory of Simple
Liquids} 2nd ed. (New York: Academic)

\bibitem{allen} Allen M P and Tildesley D J 1987 {\it Computer Simulation
of Liquids} (Oxford: Clarendon)

\bibitem{frenkelsmit} Frenkel D and Smit B 1996 {\it Understanding
                      Molecular Simulation} (London: Academic)

\bibitem{footnote} The blob picture of a star is described in
Ref.\ \onlinecite{daoud}.

\bibitem{witten} Witten T A and Pincus P A 1986 {\it Macromolecules} {\bf 19}
2509

\bibitem{foot1} Triplet forces start to become relevant only of three
spheres of diameter $\sigma$ exhibit a triple overlap within their
corona, corresponding roughly to densities $\rho > 2 \rho^{*}$. 

\bibitem{foerster} Buitenhuis J and F{\"o}rster S 1997 {\it J. Chem. Phys.}
{\bf 107} 262

\bibitem{ry} Rogers F A and Young D A 1984 {\it Phys. Rev. A} {\bf 30} 999

\bibitem{hanver} Hansen J P and Verlet L 1969 {\it Phys. Rev.} {\bf 184} 151

\bibitem{hansenschiff}  Hansen J P and Schiff D 1973 {\it Mol. Phys.}
                        {\bf 25} 1281

\bibitem{hartmut} L{\"o}wen H 1994 {\it Phys. Rep.} {\bf 237} 249

\bibitem{kang} Kang H S, Lee C S, Ree T and Ree F H 1985
               {\it J. Chem. Phys.} {\bf 82} 414

\bibitem{lutsko} Lutsko J F and Baus M 1991 {\it J. Phys.: Condens. Matter}
                 {\bf 3} 6547

\bibitem{barkerhenderson} Barker J A and Henderson D 1967
                          {\it J. Chem. Phys.} {\bf 47} 4714

\bibitem{WCA} Weeks J D, Chandler D and Anderson H C 1971
              {\it J. Chem. Phys.} {\bf 54} 5237

\bibitem{bccnote} Notice that the specific value of
   $\eta_{c}$ observed here, is a consequence of the identification
   of $\lambda$ with $a_{fcc}$, resulting in $\eta_{c}=0.74$, which is
   the maximum packing fraction for a $fcc$ crystal of hard spheres.
   If, e.g., $\lambda$ is identified with the nearest-neighbour
   distance in a $bcc$ crystal, $\eta_{c}=0.68$. Nevertheless, all
   described features of $\eta_{HS}(\eta)$ remain the same.

\bibitem{bolhuis} Bolhuis P G, Frenkel D, Mau S-C, and Huse D A 1997
                  {\it Nature} {\bf 388} 235

\bibitem{cates} Witten T A, Pincus P A and Cates M E 1986 {\it Europhys.
Lett.} {\bf 2} 137

\bibitem{uhlenbeck} Uhlenbeck G E and Ford G W 1962 in {\it Studies in Statistical
                    Mechanics} De Boer J and Uhlenbeck B E (Eds.) 
                    (Amsterdam: North-Holland)  

\bibitem{still} Stillinger F H and Stillinger D K 1997 {\it Physica A}
{\bf 244} 358 and references therein

\bibitem{pen} Likos C N, Watzlawek M and L{\"o}wen H 1998 {\it Phys. Rev. E}
              accepted for publication

\end{references}
\end{document}